\documentclass[manuscript]{aastex}

\shorttitle{Mid-infrared Sky Brightness and Fluctuation}
\shortauthors{Pyo et al.}

\usepackage{natbib}
\begin{document}

\title{Brightness and Fluctuation of the Mid-Infrared Sky from \textit{AKARI} Observations
  towards the North Ecliptic Pole}

\author{Jeonghyun~Pyo\altaffilmark{1},
  Toshio~Matsumoto\altaffilmark{2,3,4},
  Woong-Seob Jeong\altaffilmark{1},
  and
  Shuji~Matsuura\altaffilmark{3}
}

\affil{\altaffilmark{1}Korea Astronomy and Space Science Institute (KASI),
  Daejeon 305-348, Republic of Korea}
\email{jhpyo@kasi.re.kr}
\affil{\altaffilmark{2}Department of Physics and Astronomy, Seoul National
  University, Seoul 151-742, Republic of Korea}
\affil{\altaffilmark{3}Institute of Space and Astronautical Science (ISAS),
  Japan Aerospace Exploration Agency (JAXA), Kanagawa 252-5210, Japan}
\altaffiltext{4}{present address: Academia Sinica Institute of Astronomy and
  Astrophysics (ASIAA), Taipei 10617, Taiwan}

\begin{abstract}
  We present the smoothness of the mid-infrared sky from observations by
  the Japanese infrared astronomical satellite \textit{AKARI}\@.
  \textit{AKARI} monitored the north ecliptic
  pole (NEP) during its cold phase with nine wavebands covering from 2.4 to 24\,\micron\@,
  out of which six mid-infrared bands were used in this study.
  We applied power spectrum analysis to the images in order to search
  for the fluctuation of the sky brightness.
  Observed fluctuation is explained by fluctuation of photon noise, shot noise
  of faint sources, and Galactic cirrus.
  The fluctuations at a few arcminutes scales at short mid-infrared
  wavelengths (7, 9, and 11\,\micron) are largely caused by
  the diffuse Galactic light of the interstellar dust cirrus.
  At long mid-infrared wavelengths (15, 18, and 24\,\micron),
  photon noise is the dominant source of fluctuation over the scale from
  arcseconds to a few arcminutes.
  The residual fluctuation amplitude at $200\arcsec$ after removing these
  contributions is at most
  $1.04 \pm 0.23\,\mbox{nW}\,\mbox{m}^{-2}\,\mbox{sr}^{-1}$ or 0.05\% of the
  brightness at 24\,\micron{} and at least
  $0.47 \pm 0.14\,\mbox{nW}\,\mbox{m}^{-2}\,\mbox{sr}^{-1}$ or 0.02\% at
  18\,\micron\@.
  We conclude that the upper limit of the fluctuation in the zodiacal light towards the NEP is
  0.03\% of the sky brightness, taking $2\sigma$ error into account.
\end{abstract}


\section{Introduction}
\label{sec:Introduction}

Fluctuation in infrared sky brightness has been measured
for various purposes depending on the wavelength regime.
In the near-infrared, researchers have tried to discriminate the fluctuation by
the cosmic near-infrared background (CNIRB) from those by other foreground sources.
The CNIRB is the unresolved light from extragalactic sources including
the radiation from first (Population III)
stars, which reionized the Universe \citep{2002MNRAS.336.1082S}\@.
Recently, \cite{2011ApJ...742..124M} found excess fluctuation on scales
larger than $\sim 100\arcsec$ from the \textit{AKARI} north ecliptic
pole (NEP) survey at 2.4, 3.2, and 4.1\,\micron{} and interpreted it as CNIRB fluctuation.
Their results are consistent with independent fluctuation measurements from
the \textit{Spitzer} observations \citep{2005Natur.438...45K, 2007ApJ...654L...5K}\@.
In the far-infrared, the Galactic cirrus of interstellar dust is a prominent
source of diffuse brightness and fluctuation.
Far-infrared fluctuation measurements are used for studies of not only the
Galactic cirrus but also the clustering of unresolved, high-redshift objects.
The fluctuation by the cosmic far-infrared background has also been reported by several
researchers \citep[see][and references therein]{2007AdSpR..40..600J, 2011ApJ...737....2M}\@.

The zodiacal light (ZL) is a conspicuous probable source of infrared
fluctuation, because it dominates the sky brightness over
a wide wavelength range of the infrared \citep{1998A&AS..127....1L}\@.
\cite{1997A&A...328..702A} analyzed five $0\fdg5 \times 0\fdg5$ images from
25\,\micron{} observations of the Infrared Space Observatory (ISO) and concluded that
the upper limit of the ZL fluctuation is 0.2\% of brightness,
which corresponds to $\sim 5\,\mbox{nW}\,\mbox{m}^{-2}\,\mbox{sr}^{-1}$ towards
the NEP\@.
However, the ISO observation had a sparse resolution of 3\arcmin{} and
the 0.2\% fluctuation is apparently too large when compared with modern
CNIRB fluctuation measurements.

In this paper we make use of the \textit{AKARI}'s NEP monitor
observations in six mid-infrared wavebands to measure the fluctuations
in the sky brightness.
We will concentrate on searching for residual fluctuation at around
$200\arcsec$ scale after eliminating the contributions from known fluctuation
sources.
Section~\ref{sec:Observations} introduces the details of the observations and
the image reduction and calibration processes.
In Section~\ref{sec:SeasonalVariation}, we attempt sinusoidal fits to the seasonal
variation of the sky brightness, derived from the ZL\@.
Section~\ref{sec:FluctuationAnalysis} shows the results from the fluctuation
analysis and Section~\ref{sec:Discussion} discusses them.
We summarize the paper in Section~\ref{sec:Conclusion}\@.

\section{Observations and Data Reduction}
\label{sec:Observations}

\subsection{Monitor Observations}
\label{subsec:MonitorObservations}

\textit{AKARI} rotates around the Earth along a Sun-synchronous polar orbit of an altitude of
$700\,\mbox{km}$ \citep{2007PASJ...59S.369M}\@.
Owing to frequent opportunities to observe the NEP
over the whole mission period, it observed regions close to the NEP twice
a month in order to check the detectors' stability \citep{2008PASJ...60S.375T} and to
study the CNIRB \citep{2011ApJ...742..124M}\@.
The Monitor Fields were observed with the Infrared Camera (IRC) composed of
three channels, NIR, MIR-S, and MIR-L\@.
Each channel is equipped with three photometric bands.
In this paper, we will focus on the two mid-infrared channels, MIR-S and
MIR-L\@.
Table~\ref{tab:IRCChannels} lists the imaging specifications of IRC MIR-S and MIR-L channels and
bands \citep{2007PASJ...59S.401O, AKARI_IRCDUM}\@.
With six mid-infrared bands, the IRC covers a wavelength range from 7 to 24\,\micron\@.

\begin{figure*}
  \centering
  \includegraphics[width=\textwidth]{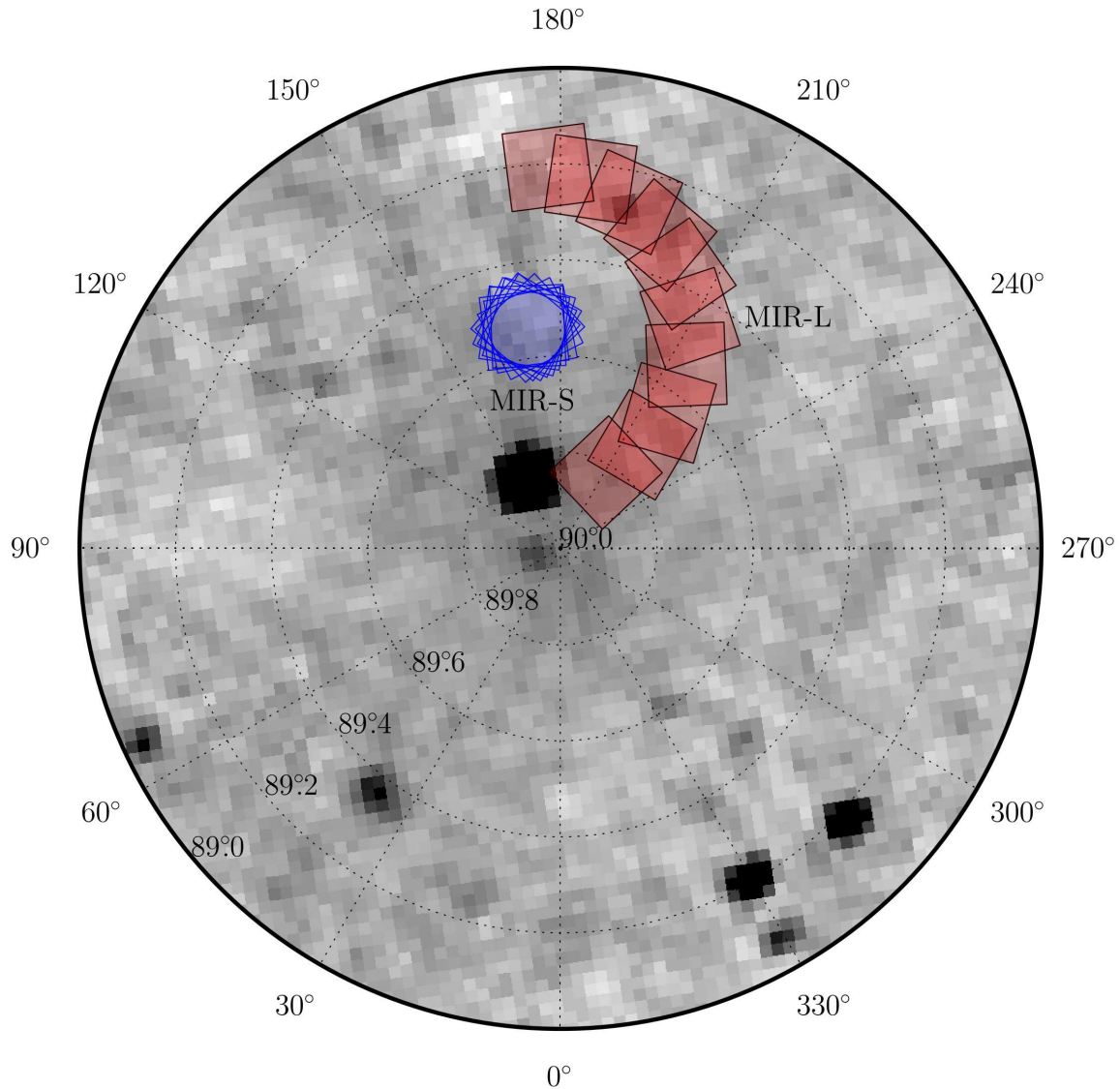}
  \caption{Monitor Fields in the ecliptic coordinates.  The blue boxes
    with light borders (colored blue) are the fields
    observed by the MIR-S channels, while red ones with dark borders
    are the fields observed by the MIR-L.  The uppermost red box
    corresponds to the MIR-L field of observation ID 5121014-001, and the ID
    increases in the clockwise direction (ref.\
    Table~\ref{tab:ObservationInfo})\@.  The background image is from the Improved
    Reprocessing of the IRAS Survey (IRIS; \citealt{2005ApJS..157..302M}) Atlas
    at 12\,\micron{} retrieved from the NASA/IPAC Infrared Science Archive.
    The center of the image is at the NEP, close to which the planetary nebula
    NGC~6543 is apparent.  See the electronic edition of the Journal for
    a color version of this figure.\label{fig:NEP}}
\end{figure*}

The monitor observation lasted from June 2006 to August 2007.
Because \textit{AKARI} suffered from the impact of scattered Earth-shine
during the northern hemisphere's summer season
\citep{AKARI_FISDUM, 2010A&A...523A..53P}, however, we excluded the observations
in that period.
Table~\ref{tab:ObservationInfo} lists the pointed observations\footnotemark{} used in this paper
and Figure~\ref{fig:NEP} shows the observed fields in the ecliptic coordinates.
\footnotetext{The observation data are available on the \textit{AKARI} homepage
(\url{http://darts.isas.jaxa.jp/astro/akari}) of the ISAS/JAXA Data Archives
and Transmission System (DARTS) by querying with the observation IDs\@.}
Note that the fields of MIR-L channel
rotated around those of the MIR-S channel
due to the different off-axis positions among the channels in the focal plane:
The field-of-view of the MIR-L channel is about $20\arcmin$ away from that of
MIR-S \citep{2007PASJ...59S.401O}\@.

\subsection{Data Reduction}
\label{subsec:DataReduction}

The observation data are reduced with the \textit{AKARI} IRC Image Data Reduction
Pipeline\footnotemark{} (version 20091022)\@.
\footnotetext{Available in the \textit{AKARI} Observers Page
  (\url{http://www.ir.isas.jaxa.jp/ASTRO-F/Observation/})\@.}
To correct the dark currents in the exposure frames, the reduction
pipeline by default subtracts the super-dark frames generated from the dark current
measurements in the Large Magellanic Cloud observations \citep{AKARI_IRCDUM}\@.
For our analysis, we modified the pipeline and subtracted
the dark frame taken just before exposures in each pointed observation,
the so-called self-dark.
This modification mitigated the effect of the dark-current variation over long-term in the
exposure frames \citep{AKARI_IRCDUM}\@.

\begin{figure*}
  \centering
  \includegraphics[width=\textwidth]{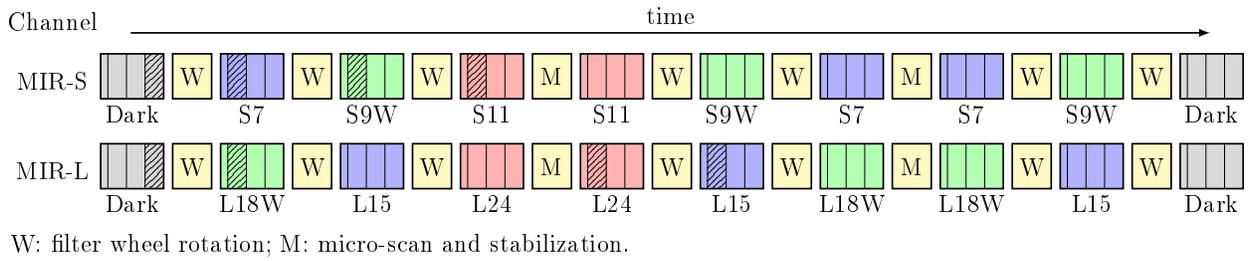}
  \caption{Sequence of an \textit{AKARI} IRC observation with the AOT IRC03 for
    the MIR-S and MIR-L channels.
    Each band of the MIR-S and MIR-L channels has two (S11 and L24 bands) or
    three (S7, S9W, L15, and L18W bands) chances of exposure, shown as the
    boxes colored blue (S7 and L15), green (S9W and L18W), or red (S11 and L15)\@.
    In each exposure, one short-exposure (0.5844 second) and three
    long-exposure (16.3632 second) frames are taken.
    The exposures are intervened by filter wheel rotations (yellow boxes with
    W) and micro-scans (yellow boxes with M) for attitude stabilization.
    Before and after the photometric exposures, dark frames (gray boxes) are
    taken.
    The frames used in this paper are indicated by hatching.
    See the electronic edition of the Journal for a color version of this
    figure.\label{fig:ObservationSequence}}
\end{figure*}

\textit{AKARI} provides IRC observers with five observation modes, the so-called
Astronomical Observation Templates (AOTs)\@.
We selected AOT IRC03, which enabled us to obtain images in nine bands with
a single pointed observation including the near-infrared channel.
The near-infrared images simultaneously taken with MIR-S and MIR-L
were separately studied by \cite{2011ApJ...742..124M}\@.
In the IRC03 template, we have two or three exposures for each mid-infrared band, each of which
consisted of one short-exposure (0.5844 second) and three
long-exposure (16.3632 second) frames (\citealt{2007PASJ...59S.401O}; see Figure~\ref{fig:ObservationSequence})\@.
We excluded short-exposure frames due to their low
signal-to-noise ratio.
On the other hand, the long-exposure (LE) frames
suffered from after-effects.
The second and third LE frames show systematically larger electron counts than
the first (LE1) does, and are ignored.
For the MIR-S bands, the LE1 frames in later exposures are also ignored, because
they are systematically brighter than the one in the first exposure.
In the case of the MIR-L bands, the situation is more complicated.
The L18W band comes first in the sequence of the MIR-L channel exposures and the
electron count is the largest in that band.
The first L18W-band exposure significantly affects the following first L15- and
L24-band exposures, which record higher counts than the second ones do.
Therefore, we took the LE1 frames in the second exposures of the L15 and L24 bands,
while the LE1 frame in the first exposure was chosen for the L18W band.
Consequently, only one image of $\sim 16$-second exposure was taken for each band from every pointed
observation.
Therefore the images used in this study are about $\sqrt{6}$ or $3$
times shallower than those obtained through the coadding procedure of the reduction
pipeline.

\begin{figure*}
  \centering
  \includegraphics[scale=0.6916]{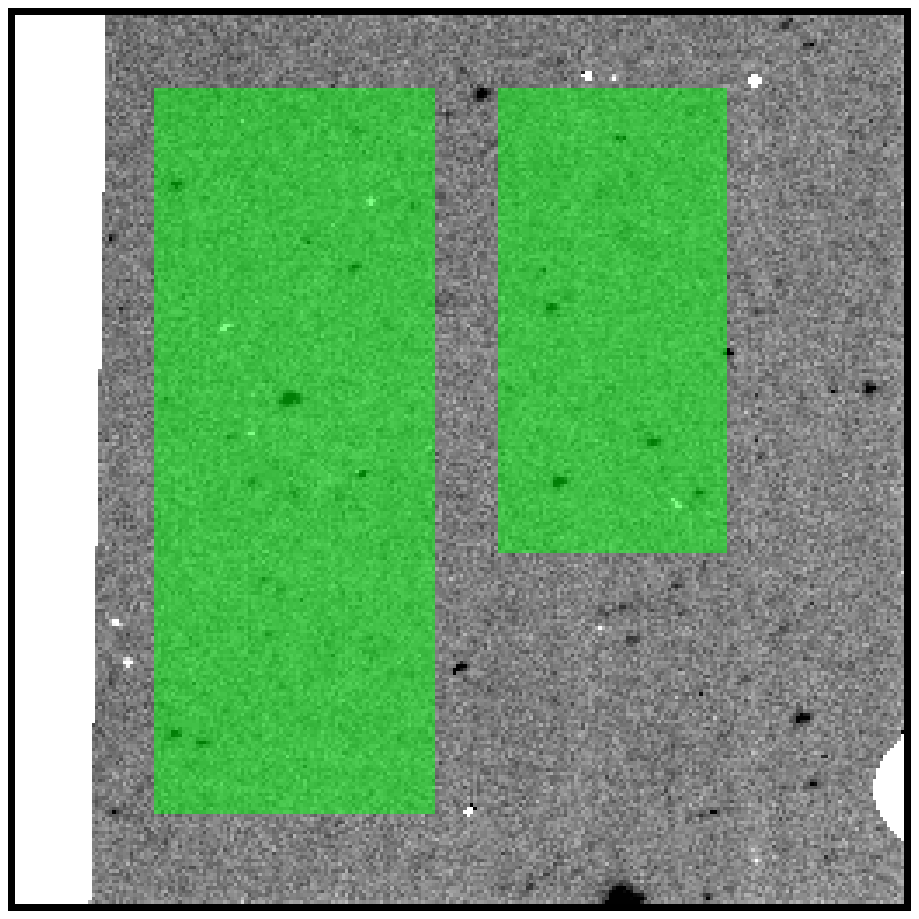}\qquad
  \includegraphics[scale=0.7421]{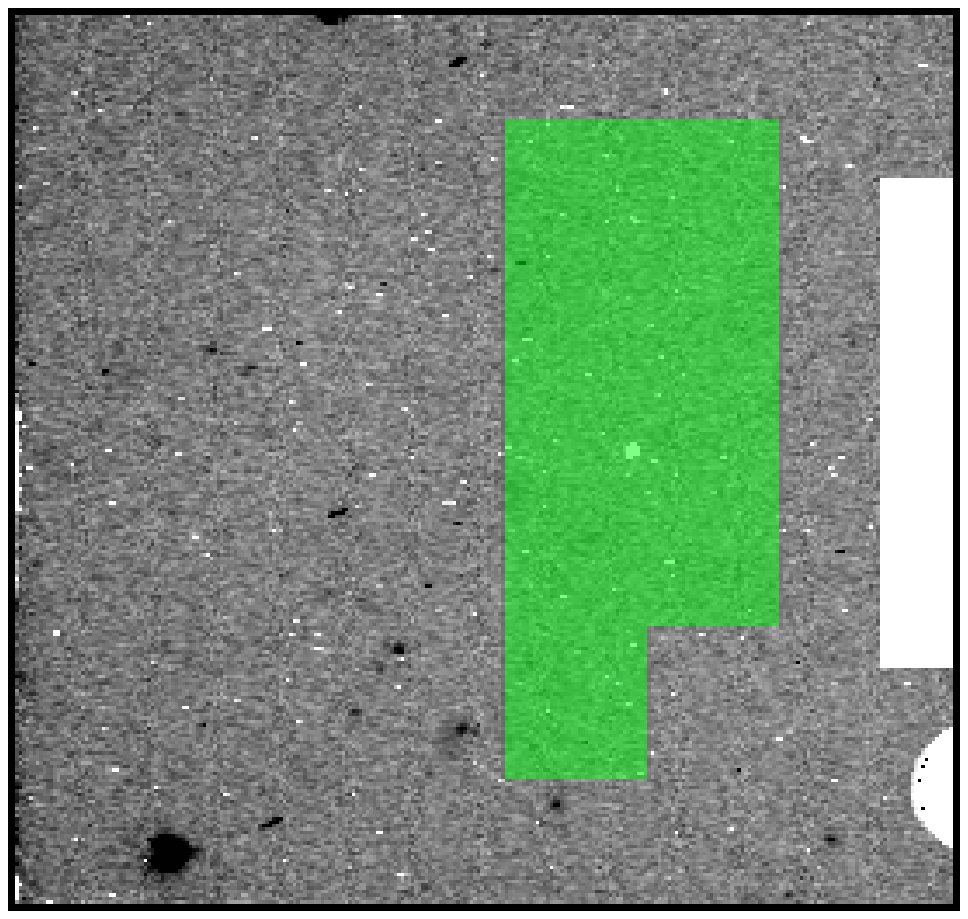}
  \caption{Regions (green-color areas) used in the analysis for the MIR-S (left)
  and MIR-L (right) channels.  The background images are the real
  observations in the S9W and L18W bands after
  processing with the \textit{AKARI} IRC image reduction pipeline for the observation
  ID 5121016-001\@.  The images are inverted for visibility.  Blank regions on
  the left (MIR-S), and right (MIR-L) sides in the frames are
  masked to measure the dark current.  Small, circular areas on the
  right-bottom side are masked to avoid
  the emission from internal lamps.  The sizes of images are scaled to
  match their angular scales.
  See the electronic edition of the Journal for a color version of this
  figure\@.\label{fig:Masks}}
\end{figure*}

Parts of images affected by
internal lamps and bright lines of pixels were masked out.
Additionally, we masked the left half of the MIR-L channel frames, in which a
pattern of internally scattered light was significant.
The regions used in the analysis are shown in Figure~\ref{fig:Masks}\@.

The stars in the images were removed by masking the pixels whose values were
larger than or smaller than three times the standard deviation with respect to the
average, along with $3 \times 3$ pixels surrounding them.
The masking was repeated until there remained no more pixels to be clipped.
The fractions of the masked pixels were $\lesssim 6\%$ for the MIR-S bands, $\sim 9\%$
for the L15 and L18W bands, and $\sim 11\%$ for the L24 band.
We did not combine the reduced images to investigate the seasonal variation of the
brightness.

The dark frames were processed in the same way as the exposure frames except
that we skipped the dark subtraction procedure.
After the whole reduction processes, we obtained the average image of the dark
frames for each channel and subtracted it from individual dark frames to strip the
large-scale patterns in the dark current.
The patterns were especially significant in the MIR-S channel dark frames and
had an impact on the fluctuation measurements on scales larger than about
$50\arcsec$\@.
These large-scale patterns were suppressed in the exposure frames when
subtracting the dark frames from them pixel by pixel.

\subsection{Absolute Calibration}
\label{subsec:AbsoluteCalibration}

We used the observations of the Diffuse
Infrared Background Experiment (DIRBE) on-board the \textit{Cosmic Background Explorer} (\textit{COBE}) as a
reference for the absolute calibration of the diffuse light.
The method is similar to that used for the calibration of the \textit{AKARI} IRC All-Sky
Survey \citep{2010A&A...523A..53P}\@.
If we assume that the sky brightness at mid-infrared wavelengths follows the
blackbody spectrum \citep{2002A&A...393.1073L},
the spectral energy distribution (SED) of the sky
brightness $I_{\nu}(\lambda)$ can be written as
\begin{equation}
  I_{\nu}(\lambda) = \tau B_{\nu}(\lambda,\,T),
  \label{eqn:SED1}
\end{equation}
where $\tau$ and $T$ are the optical depth and the color temperature,
respectively, and $B_{\nu}(\lambda,\,T)$ is the Planck function at the
wavelength $\lambda$\@.
The subscript $\nu$ in $I_{\nu}$ and $B_{\nu}$ means that the unit of
brightness is given in the per-frequency unit.
Use of a per-frequency unit is not a mandatory rule, but is done to follow the
convention of adopting the unit of MJy in the previous ZL studies.
The assumption of wavelength-independent optical depth and color temperature is
acceptable for the ZL-dominant background at the mid-infrared wavelengths
\citep{1998A&AS..127....1L}\@.

Our model SED, Equation~(\ref{eqn:SED1}), is a simple, empirical model and
not a physically meaningful one.
A physical model has to include i) the ZL,
ii) the integrated star
light, iii) the diffuse Galactic light, and iv) the extragalactic background
light \citep{1998A&AS..127....1L}\@.
For the absolute calibration, however, a detailed model of light sources is not
necessary and an empirical shape of SED suffices.
Thus the temperature $T$ in Equation~(\ref{eqn:SED1}) is not related
to the temperature of the interplanetary dust (IPD), or of any other sources.
On the other hand, \cite{1998EP&S...50..507O, 2000AdSpR..25.2163O} and
\cite{2009ASPC..418...29H} modeled the infrared ZL spectra with the brightness
integrals and suggested a two-temperature model of the IPD cloud in which the
hot-dust component has a temperature of about 300\,K or higher at 1\,AU and
the cold-dust component has a temperature of about 266\,K or lower.
But the hot-dust component is required to explain the part of the spectra at
wavelengths shorter than $\sim 6\,\micron$ and a single component model is
sufficient at longer wavelengths \citep{2003Icar..164..384R}\@.

We have two unknown parameters, $\tau$ and $T$, in Equation~(\ref{eqn:SED1})\@.
To fix these two parameters, the brightnesses at two wavelengths are sufficient;
\cite{2010A&A...523A..53P} used the DIRBE 4.9 and 12\,\micron{} bands to calibrate the
\textit{AKARI} All-Sky Survey at 9\,\micron{}\@.
In our case, we made use of the DIRBE 4.9, 12, and 25\,\micron{} bands, because
the MIR-S and MIR-L channels cover the wavelength range from 7 to
24\,\micron{}\@.

The DIRBE brightnesses were taken from the Calibrated Individual Observations
(CIO) dataset\footnotemark{}\@.
\footnotetext{Available in the Legacy Archive for Microwave Background Data
Analysis (LAMBDA) homepage (\url{http://lambda.gsfc.nasa.gov}) of the 
NASA\@.}
For each \textit{AKARI} MIR-S and MIR-L observation, we collected such DIRBE
observations that satisfied the following conditions: i) The difference in the
Earth's heliocentric ecliptic longitude between DIRBE and \textit{AKARI} observations is
less than $5\degr$ and ii) the separation between the observing coordinates
of the two observations is less than $\sim 10\arcmin$, half of the spatial resolution in the CIO\@.
We collected the DIRBE observations for the MIR-S and MIR-L channels independently,
because the lines-of-sight of the two channels are separated from each other (Figure~\ref{fig:NEP})\@.
The mission period of the \textit{COBE}/DIRBE was shorter than 1 year and the DIRBE
observations were not available for five \textit{AKARI} observations
of IDs from 5121016-001 to 5121020-001\@.
For each \textit{AKARI} observation, about 20 sets of DIRBE brightnesses were selected.
We took the average of the selected brightnesses.

\begin{figure*}
  \centering
  \includegraphics[width=0.49\linewidth]{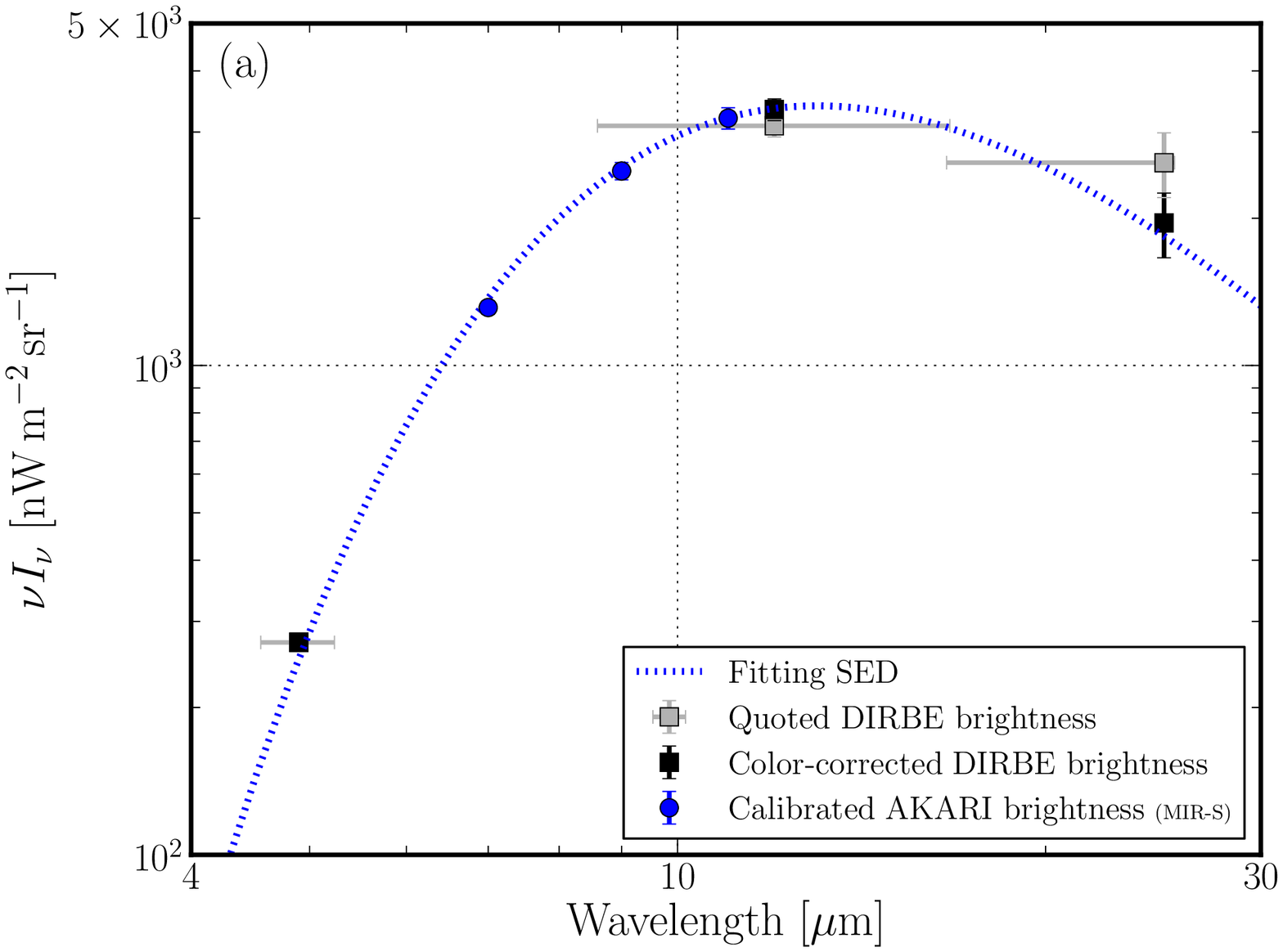}
  \includegraphics[width=0.49\linewidth]{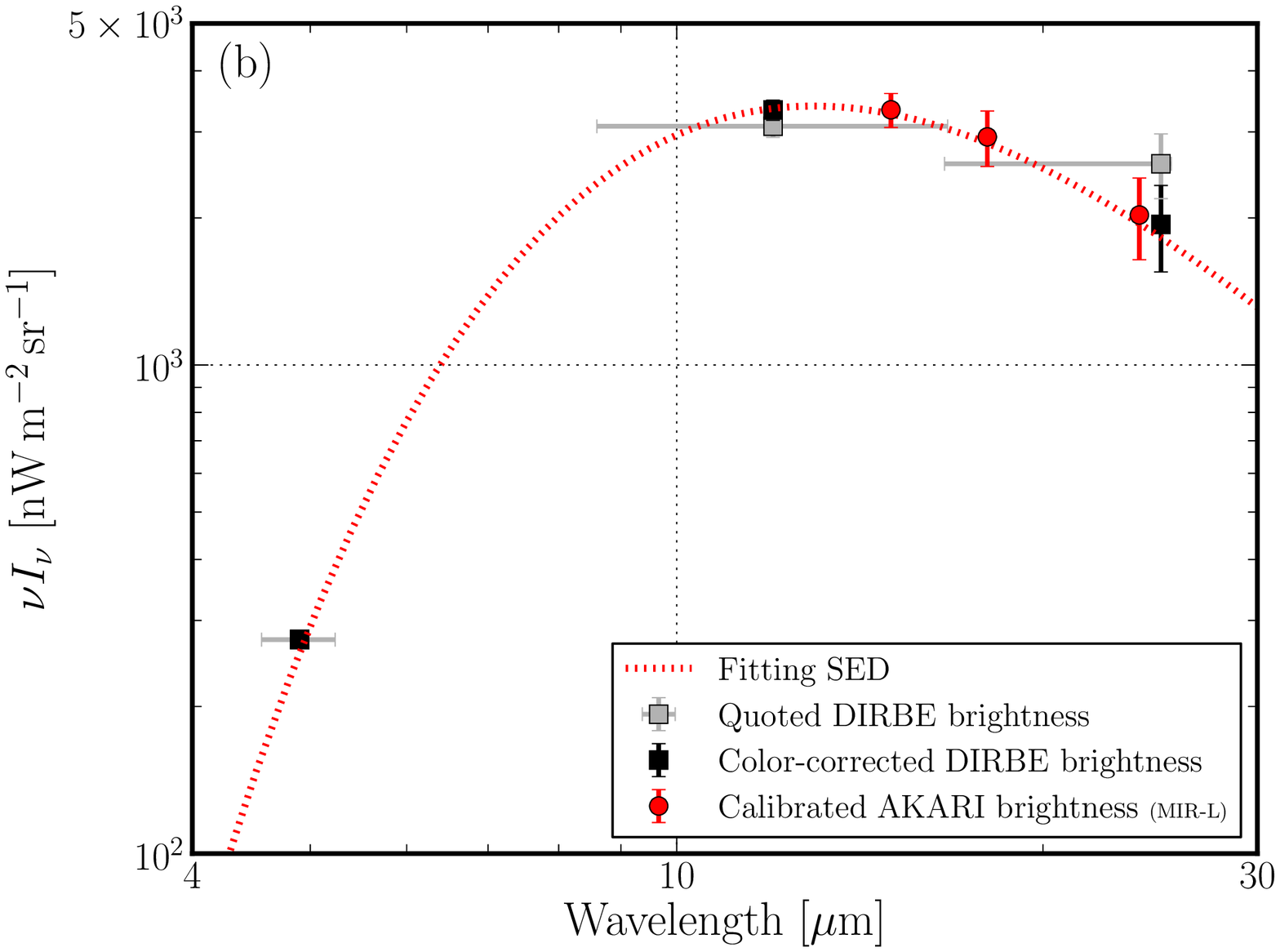}
  \caption{Calibration process for the MIR-S (a) and MIR-L bands (b).
    Data are plotted for the observation ID 5121021-001\@.
    Gray squares are the quoted DIRBE brightnesses at 4.9, 12, and
    25\,\micron{} bands, corresponding to the \textit{AKARI} observation configuration.
    The horizontal bars are the full-widths at half-maximum (FWHM) of the
    spectral response curves of the DIRBE bands, while the vertical bars are the
    errors in the DIRBE brightnesses due to the gain uncertainty
    \citep{1998ApJ...508...25H}\@.
    The black squares are the color-corrected DIRBE brightnesses based on the SED
    drawn with green (a) or red (b) dotted line.
    The SED is determined by fitting a blackbody SED to the quoted DIRBE
    brightnesses, while applying color-correction.
    Note that the SED curves and the DIRBE brightnesses in (a) and (b) are
    slightly different from each other, because the observing coordinates of the MIR-S and
    MIR-L channels do not coincide.
    Green (a) and red (b) circles are the calibrated \textit{AKARI} brightnesses in the
    MIR-S and MIR-L bands, respectively.
    See the electronic edition of the Journal for a color version of this
    figure\@.\label{fig:CalibrationSED}}
\end{figure*}

The DIRBE CIO dataset provides the quoted brightness in which the calibration
assumed that the source spectrum follows $I_{\nu} \propto 1 / \nu$\@.
Because we assume a blackbody SED,
we made the color correction on the DIRBE brightnesses, which depends on temperature only.
For each set of three DIRBE brightnesses observed simultaneously, we determined the temperature
$T$ and the optical depth $\tau$ with the least-squares fit using the following
model:
\begin{equation}
  I_{\nu}^{i} = \tau B_{\nu}(\lambda^{i},\,T) K^{i}(T),
\end{equation}
where the superscript $i$ runs over three DIRBE bands and $I_{\nu}^{i}$,
$\lambda^{i}$, and $K^{i}$ are the quoted brightness, the effective wavelength,
and the color-correction factor for the $i$-th band, respectively.
The equation for calculation of the color-correction factor is given in
\cite{Hauser:DIRBEExpl}\@.
The quoted and color-corrected DIRBE brightnesses and the fitting blackbody SED
corresponding to the \textit{AKARI}'s observation ID 5121021-001
are shown in Figure~\ref{fig:CalibrationSED} for instance.

Once the temperature and optical depth were determined, we calculated the sky
brightnesses at the MIR-S and MIR-L bands' reference wavelengths with
Equation~(\ref{eqn:SED1})\@.
The calculated brightnesses were divided by the ADU counts from the \textit{AKARI} observations.
Consequently, we obtained seven (MIR-S) or five (MIR-L) ratios between the calculated and measured
brightnesses for each band and took their averages as the calibration factors.
In Figure~\ref{fig:CalibrationSED}, the calibrated \textit{AKARI} brightnesses are shown
with green (for MIR-S channel) or red (for MIR-L channel) circles.
The calibration factors used in this work are listed in Table~\ref{tab:IRCChannels}\@.

The calibration factors for diffuse sky brightness are smaller than those for the point sources
\citep{2008PASJ...60S.375T} by factors of about 0.85--0.89 for MIR-S bands and about
0.56--0.78 for MIR-L bands.
The ratios between the diffuse and point source calibration factors are listed
in Table~\ref{tab:IRCChannels}\@.
The difference of calibration factors between the point and the diffuse sources is
due to i) the point spread function (PSF) cut-off in the aperture photometry
of the point sources  and ii) the internal scattering and diffraction of light within a
detector array \citep{2011PASP..123..981A}\@.
The former causes the point source calibration factor to be overestimated, but does
not cause a problem in point source photometry if a consistent aperture size is used for
both the calibration and the photometry.
On the other hand, the latter decreases the diffuse source calibration factor
with respect to the ``ideal'' absolute calibration.
As long as the internal scattering and diffraction are proportional to the incident brightness,
our calibration method correctly takes them into account.

We note that our calibration factors are subject to the uncertainty in the
absolute gain calibration of DIRBE, 3.0\% at 4.9\,\micron{}, 5.1\% at
12\,\micron{}, and 15.1\% at 25\,\micron{} \citep{1998ApJ...508...25H}\@.
The uncertainties induce $\sim 10\%$ errors in our calibration.
However, we will ignore those systematic errors because our central concern is
to measure the fluctuation with respect to the sky brightness.

\section{Seasonal Variation of the Sky Brightness}
\label{sec:SeasonalVariation}

\begin{figure*}
  \centering
  \includegraphics[width=0.49\textwidth]{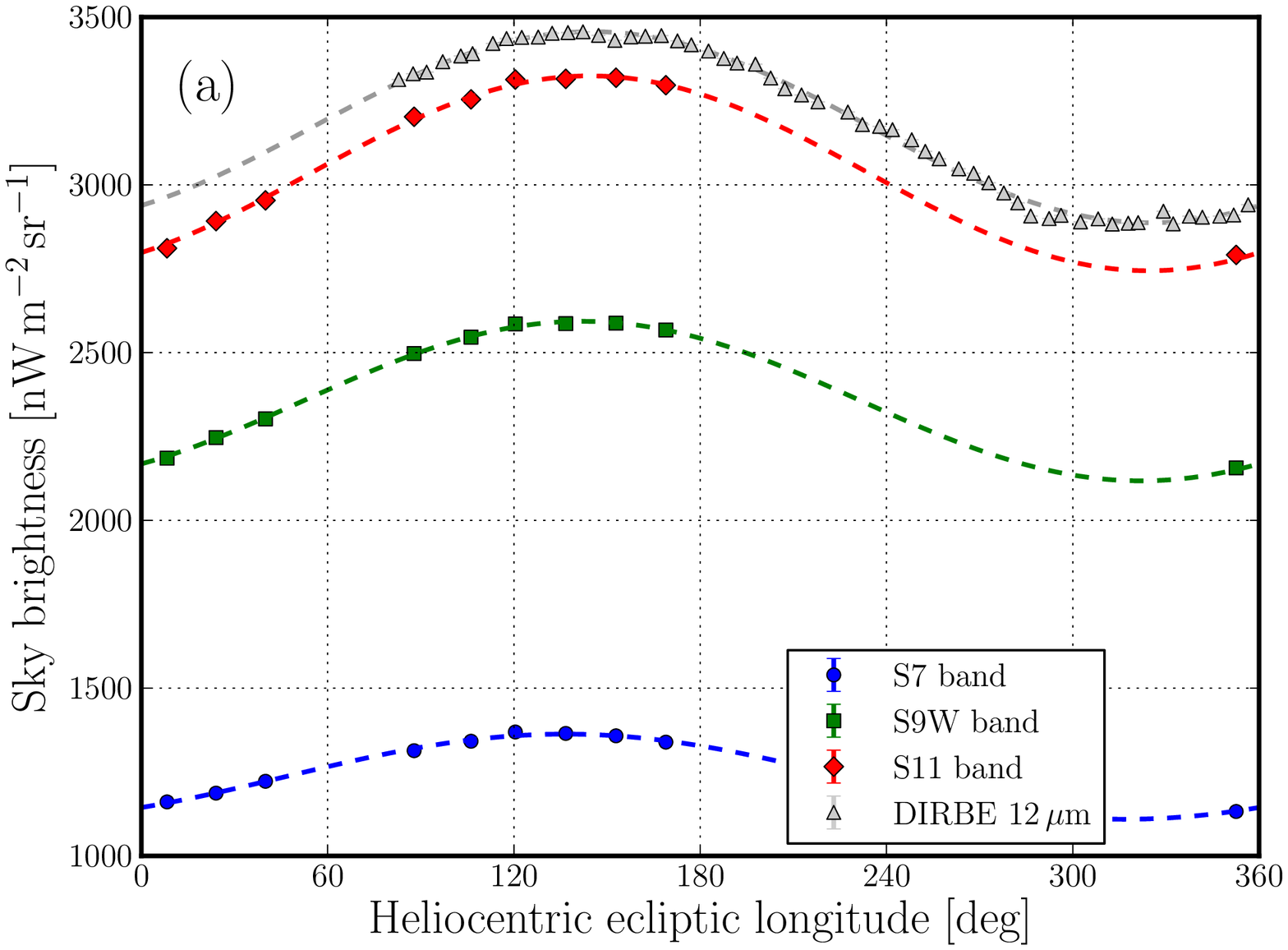}
  \includegraphics[width=0.49\textwidth]{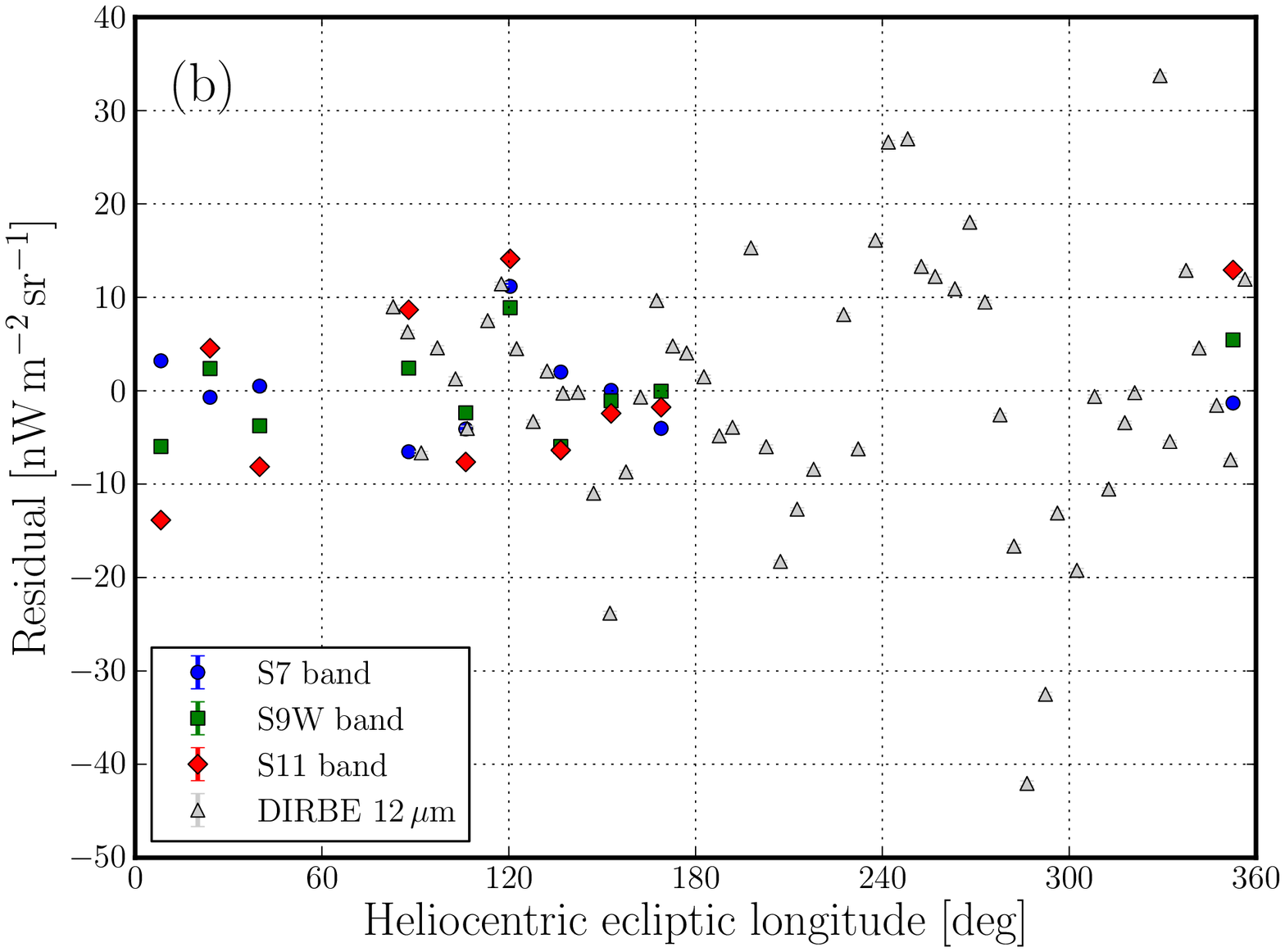}
  \\[2ex]
  \includegraphics[width=0.49\textwidth]{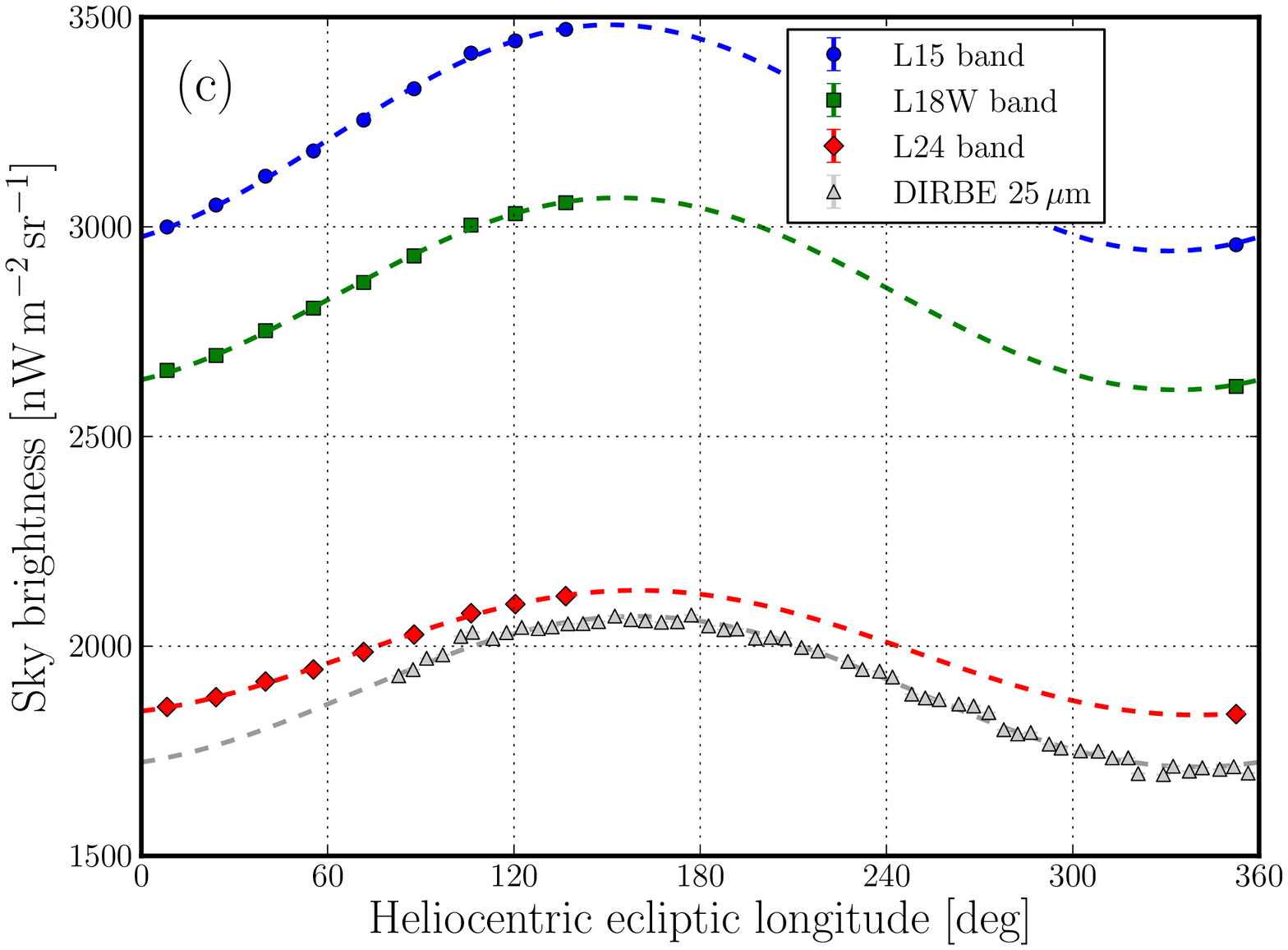}
  \includegraphics[width=0.49\textwidth]{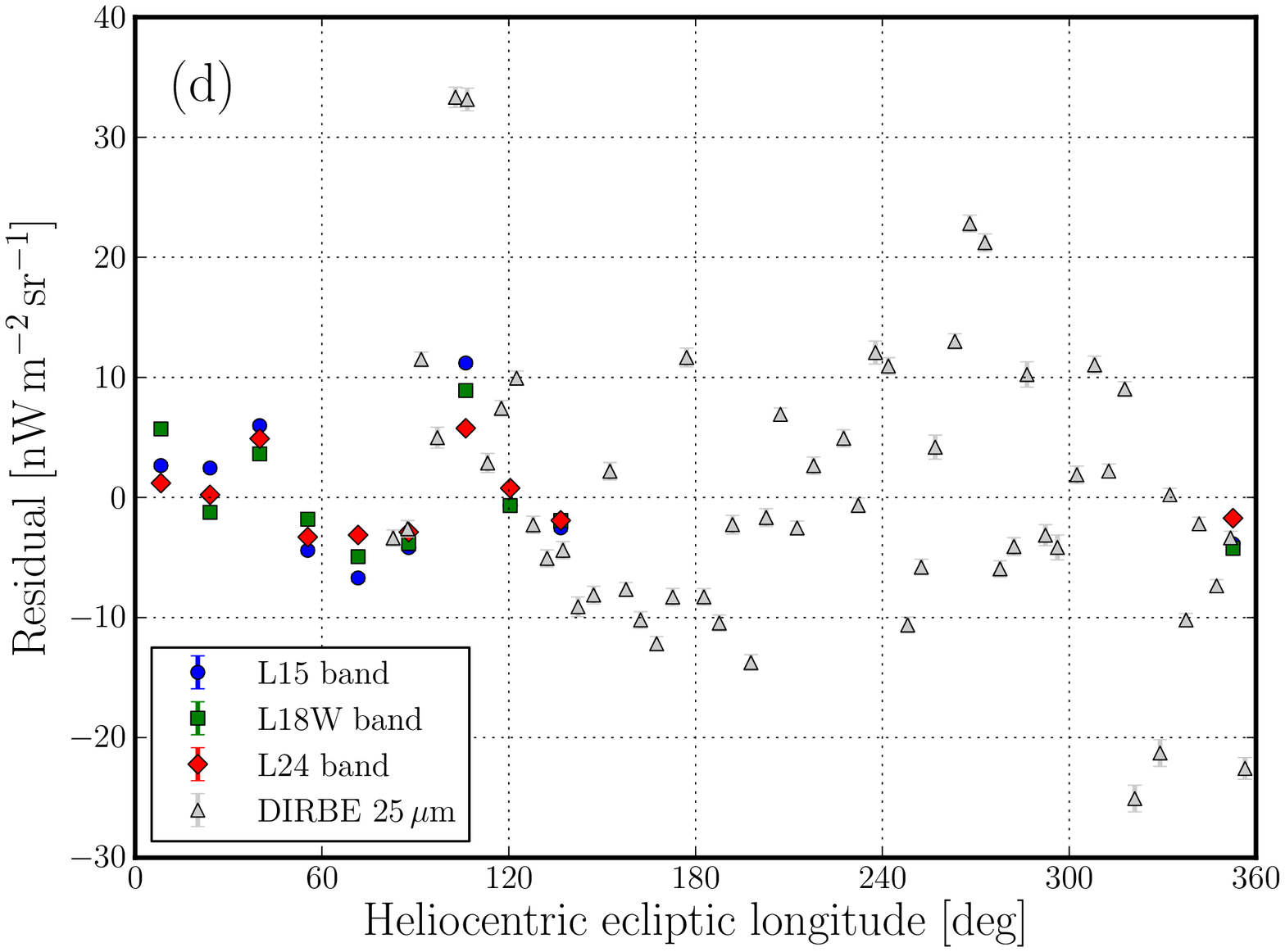}
  \caption{Sinusoidal fittings to the sky brightnesses observed by the
    \textit{AKARI} NEP Monitor Observations (blue circles, green squares, and red diamonds) and
    the DIRBE observations (gray triangles)\@.  Panels (a) and (c) in the left column
    show the observed sky brightnesses (symbols) as functions of the Earth's
    heliocentric ecliptic longitude and sine curves (dashed lines) fitting to the
    symbols.  Panels (b) and (d) in the right column show the residuals
    after subtracting the fitting curves from the brightnesses.
    In each panel, we simultaneously
    plot the results for three bands of MIR-S and MIR-L channels in the top
    and bottom rows, respectively.
    The results for the DIRBE 12 and 25\,\micron{} are shown
    in the top and bottom rows, respectively, and drawn with gray points and lines.
    The error bar is drawn for each symbol, but
    it is too small to be apparent.
    See the electronic edition of the Journal for a color version of this
    figure. \label{fig:SineFitting}}
\end{figure*}

The sky brightness observed on the Earth-bound orbit changes with time due to
the annual variation of the ZL
\citep{1988A&A...196..277D, 1988ApJ...335..468R, 1995ApJ...455..677V,
1998ApJ...508...44K, 1998EP&S...50..501K, 2010A&A...523A..53P}\@.
For detailed studies of the Solar system's IPD cloud, such full
models as those of \cite{1998ApJ...496....1W} and
\cite{1998ApJ...508...44K} are required.
However, complete modeling of the IPD cloud is beyond the scope of this work.
Instead, we tried simple sinusoidal fitting to the observed sky brightnesses,
which assumes that the ZL at high ecliptic latitude is dominated by the smooth
cloud component of the IPD cloud \citep{1998ApJ...508...44K}\@.
The fitting function is given as a function of the Earth's heliocentric
ecliptic longitude, $\lambda_{\earth}$, at the epoch of observation:
\begin{equation}
  I_{\nu}(\lambda_{\earth}) = a \sin \left( \lambda_{\earth} - b \right) + c,
\end{equation}
where $a$, $b$, and $c$ are the amplitude, phase, and
average brightness of the annual variation, respectively.
In the fitting process, we fixed the period of the sine curve at $360\degr$, which is
a reasonable assumption.
The results are plotted in Figure~\ref{fig:SineFitting} and
arranged in Table~\ref{tab:SineFitting}\@.
For comparison, the results for the DIRBE 12 and 25\,\micron{} bands
observations are shown.
The DIRBE NEP brightnesses are retrieved from the CIO dataset by collecting
the observations pointing at an ecliptic latitude larger than $89\fdg5$ while
avoiding the region within $0\fdg5$ of NGC~6543\@.
The brightnesses within $5\degr$ intervals of the Earth's heliocentric ecliptic
longitude are averaged and shown in Figure~\ref{fig:SineFitting} with gray
triangles.
Error bars are the statistical errors of the averages.

As can be seen in Figure~\ref{fig:SineFitting} and in Table~\ref{tab:SineFitting}, the
sine curves well describe the seasonal variation of the NEP
brightnesses.
The residuals after subtracting the fitting sine curves from the observed
brightnesses are within $\sim 0.4\%$ of the average brightnesses for the
\textit{AKARI} observations.
We plotted the residuals with respect to the Earth's heliocentric ecliptic longitude, as
shown in the right panels of Figure~\ref{fig:SineFitting}\@.
A clear dependency of the residuals on the longitude was not found.
Meantime, the residuals of the bands in the same channel are correlated to each
other, but no correlation is found between different channels.
Therefore, we conclude that the residuals are caused by an instrumental effect,
but not by the temporal variation of the sky brightness.
At least for the MIR-L bands, it is arguable that the residuals are not related
to the dark current, because the residuals from three bands are consistent with
each other in the sense of the ratio relative to the average brightnesses.
In the current analysis, it is difficult to clearly pinpoint the cause of the
residuals.
The observed brightnesses and the fitting parameters in the S11 and L24 bands are,
respectively, consistent with those in the DIRBE 12 and 25\,\micron{}\@.

\begin{figure}
  \centering
  \includegraphics[width=0.95\linewidth]{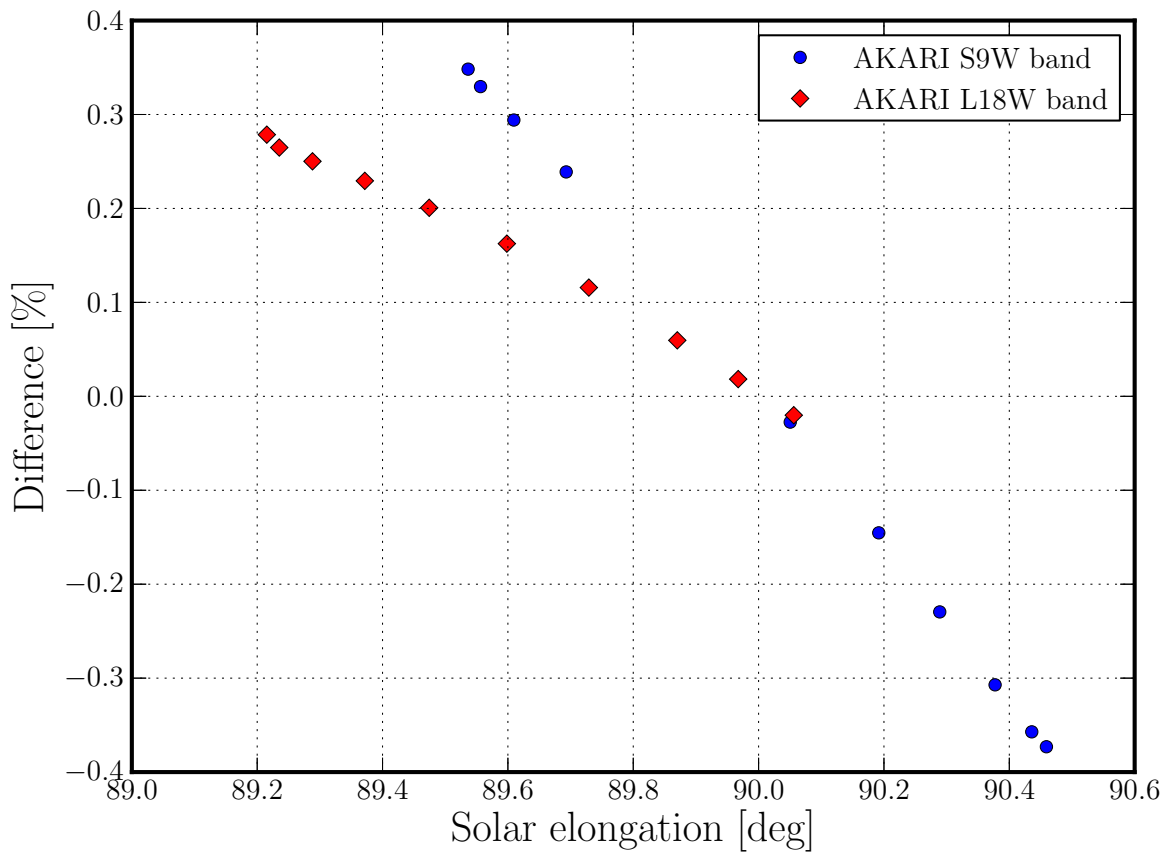}
  \caption{Difference in the ZL brightnesses between the \textit{AKARI} NEP Monitor
    Fields shown in Figure~\ref{fig:NEP} and the NEP as a function of
    solar elongation.
    The differences are calculated by subtracting the
    brightnesses at the NEP from those at the Monitor Fields, and then dividing by the
    NEP brightnesses.
    The IPD cloud model of \cite{1998ApJ...508...44K} is evaluated to
    calculate the ZL brightness.
    We plot the differences in
    the S9W (blue circles) and the L18W (red diamonds) bands.
    See the electronic edition of the Journal for a color version of this
    figure. \label{fig:NEPMonDiff}}
\end{figure}

Our observation fields deviated slightly from the exact
position of the NEP, as shown in Figure~\ref{fig:NEP}\@.
By the means of the IPD cloud model of \cite{1998ApJ...508...44K},
we examined how much difference in brightness between the NEP and the \textit{AKARI}
Monitor Fields is introduced by the deviation.
The model was evaluated for the NEP and the \textit{AKARI} Monitor Fields at the observation
epoch for each observation ID listed in Table~\ref{tab:ObservationInfo}, and then
the difference between the two brightnesses was obtained.
We calculated the model at the wavelengths of the two IRC bands, S9W from MIR-S and
L18W from MIR-L\@.
To evaluate the model at the wavelengths of $9$ and $18\,\micron$,
we turned off the inverse color-correction required to obtain the DIRBE quoted
brightness.
The emissivity modification factors at those wavelengths were obtained by
logarithmically interpolating the values given in \cite{1998ApJ...508...44K}\@.
The Earth's heliocentric coordinates required for the model calculation were retrieved from the
HORIZONS ephemeris computation service operated by the Jet Propulsion Laboratory (JPL)\@.
The calculation results are shown in Figure~\ref{fig:NEPMonDiff}\@.

The model calculation shows that the difference changes with time and is
coupled with the solar elongation, one of the
parameters most relevant to the ZL brightness.
Though the observing coordinates of the MIR-S channel are fixed at a
celestial coordinates, the solar elongation slightly changes because the
Earth revolves around the Sun.
As can be seen in Figure~\ref{fig:NEPMonDiff}, we have brightness dimmer than that of the NEP
if the solar elongation is larger than $90\degr$, and vice versa.
In the case of the MIR-S channel, the difference becomes almost zero for
the observation ID 5121021-001, at which the solar elongation is closest to
$90\degr$\@.
For this channel, the brightness difference relative to the NEP is always
smaller than or comparable to the standard deviation of the sinusoidal fitting
residuals (Table~\ref{tab:SineFitting})\@.
For the MIR-L channel, the moving line-of-sight in the celestial coordinates
introduces another modulation to the solar elongation, in addition to that caused by the Earth's motion.
That results in solar-elongation deviation from $90\degr$ larger than that for the
MIR-S channel; also, during the \textit{AKARI} observation, the MIR-L channel fields
maintain a solar elongation smaller than $90\degr$\@.
The sky brightnesses in the MIR-L fields are brighter than
those at the NEP, but the difference is smaller than two times
the standard deviation of the residuals after the sinusoidal fitting.
Consequently, though the \textit{AKARI} NEP Monitor observations were not
pointed at the exact NEP coordinates, the observed brightnesses are almost the
same as those of the NEP's within 0.4\%\@.

\section{Fluctuation Analysis}
\label{sec:FluctuationAnalysis}

We examined the fluctuation in the sky brightnesses observed in the six mid-infrared
wavebands.
To measure the fluctuation, we adopted the power spectrum method.
We first subtracted the average pixel value from every valid pixel
and filled masked pixels with zeros.
We call this the fluctuation image, $\delta I(\mathbf{x})$, where $\mathbf{x}$
is the pixel coordinates in units of radians.
Then, the two-dimensional power spectrum image $P(\mathbf{k})$ is given by 
$P(\mathbf{k}) = |\delta I_{k}(\mathbf{k})|^2$, where $\mathbf{k}$ is the
two-dimensional wavenumber vector, and the fluctuation image's
Fourier transform $\delta I_{k}(\mathbf{k})$ is defined as
\begin{equation}
  \delta I_{k}(\mathbf{k}) = \frac{1}{L_{x}L_{y}} \int \delta I(\mathbf{x})
    \exp(- i \mathbf{x}\cdot\mathbf{k})\,d^{2}\mathbf{x} .
  \label{eqn:FT}
\end{equation}
In Equation~(\ref{eqn:FT}), $L_{x}$ and $L_{y}$ are the angular sizes of the image
along $x$- and $y$-directions, respectively.
To get the one-dimensional power spectrum $P(k)$, we took the average of
$P(\mathbf{k})$ in which $\mathbf{k}$ satisfies
$k - \frac{1}{2} \Delta k \le |\mathbf{k}| < k + \frac{1}{2} \Delta k$
after masking the pixel lines, $k_{x} = 0$ and $k_{y} = 0$\@.
With a one-dimensional power spectrum, the fluctuation $F(k)$ at wavenumber
$k$ or, correspondingly, at scale $2 \pi / k$ is given by
\citep{2007ApJ...657..669T}
\begin{equation}
  F^{2}(k) = \frac{L_{x} L_{y}}{(2 \pi)^2} 2 \pi k^2 P(k) .
\end{equation}
This method is usually used in the CNIRB studies
\citep[e.g.,][]{2005Natur.438...45K, 2007ApJ...657..669T, 2011ApJ...742..124M}\@.

We applied the method to the individual images and obtained 10 spectra
for each of the MIR-S and MIR-L bands.
The same method was applied to the dark frames.
The fluctuation spectra of the sky brightness were obtained by quadratically
subtracting the dark-frame spectra from those of
exposure images.
The sky fluctuation spectra are averaged
after rejecting the largest and smallest values scale-wise, and are shown in
Figure~\ref{fig:FluctuationSpectra}\@.

\section{Discussion}
\label{sec:Discussion}

\begin{figure*}
  \centering
  \includegraphics[width=\linewidth]{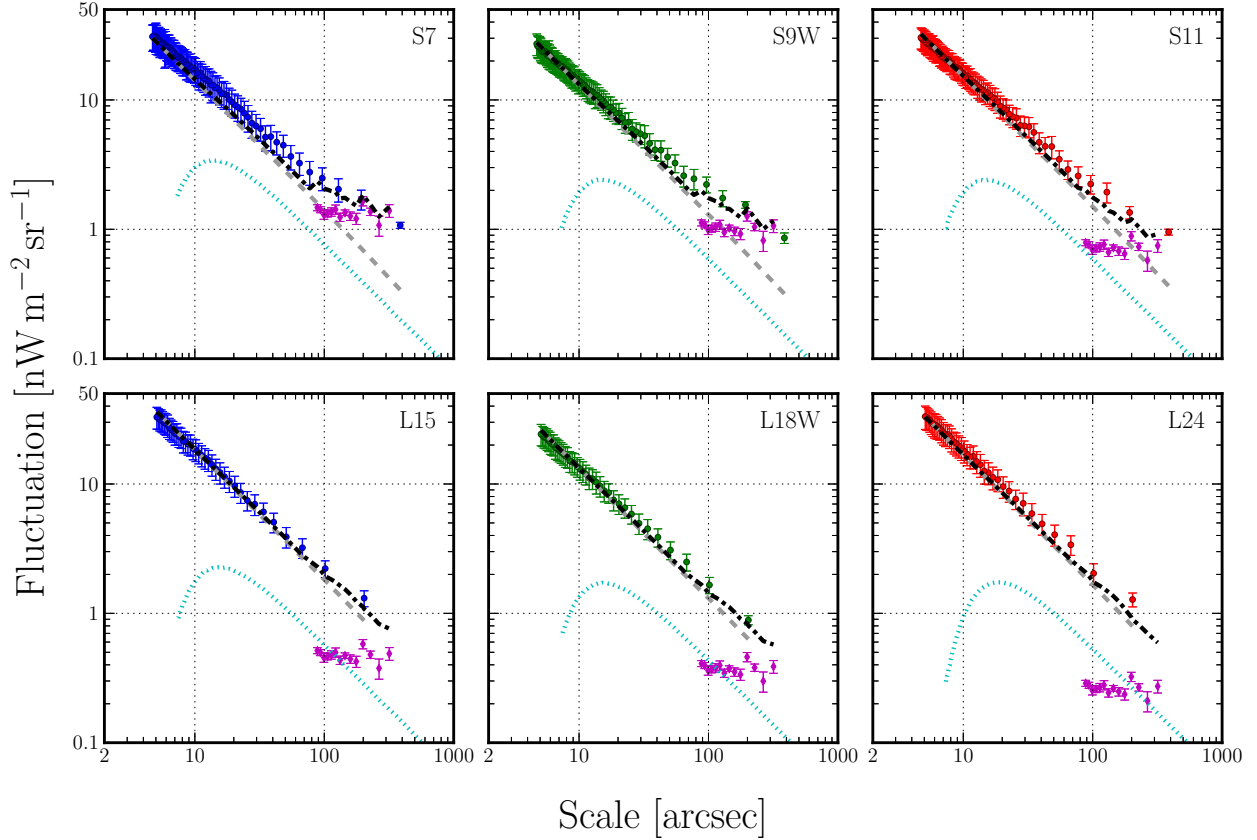}
  \caption{Fluctuation spectra of the mid-infrared sky brightness (colored circles),
    the photon noise (gray dashed lines), the Galactic cirrus (magenta
    diamonds), and the shot noise by unresolved faint sources (cyan dotted lines)\@.
    The black dash-dotted lines are the spectra obtained by quadratically
    summing the fluctuation spectra of the photon and shot noises and cirrus.
    The error bars correspond to the statistical errors of the averages.
    See the electronic edition of the Journal for a color version of this
    figure. \label{fig:FluctuationSpectra}}
\end{figure*}

At mid-infrared wavelengths the sky is very bright due to the ZL
and the electron
counts recorded in the IRC detector pixels are copious, usually larger than
two thousands.
Thus the photon noise is one of the major sources of fluctuation.
To estimate the photon-noise contribution to the fluctuation, we artificially
generated a pseudo-image filled with the values from a Poisson distribution
whose mean is fixed at the average electron count of an exposure image.
The electron count is calculated by multiplying ADU values by the instrument gain of
$6.4363\,\mathrm{e}^{-}\,\mathrm{ADU}^{-1}$\@.
We generated 100 pseudo-images for each image, calculated the
fluctuation spectra of the pseudo-images, and took the average over 100 spectra.
The photon-noise fluctuation spectrum is individually calculated for each exposure image
because the sky brightness relevant to the average electron count changes with time, as
discussed in Section~\ref{sec:SeasonalVariation}\@.
The average of 10 fluctuation spectra of the photon noise for each band is
shown in Figure~\ref{fig:FluctuationSpectra} with a gray line.

One of the sky brightness components that may explain the fluctuation is the
Galactic cirrus emission by interstellar dust.
The fluctuation spectrum of the Galactic cirrus is known to have a power-law index
in a range from 2.5 to 3.1 \citep{2005MNRAS.357..535J,
2007A&A...469..595M}\@.
To evaluate the real fluctuation spectrum of the cirrus, we used the \textit{AKARI} Far-Infrared Surveyor (FIS)
observation of the NEP field \citep{2011ApJ...737....2M} at the wavelength $90\,\micron$\@.
After removing the point sources, we measured the fluctuation of the image.
To exclude the shot noise due to faint galaxies and the effect of
large pixel size ($\sim 30\arcsec$), we used fluctuations at scales larger than
$85\arcsec$\@.
Assuming that the structure of the Galactic cirrus is the same in other wavelengths, we
converted the fluctuation measured at $90\,\micron$ to the fluctuations in the
mid-infrared bands by using the SED of the Galactic cirrus.
Regarding the cirrus SED, we adopted the recent results from
the \textit{Herschel} and \textit{Spitzer Space Telescope} observations by
\cite{2010ApJ...724L..44C}\@.
The cirrus fluctuation spectra are plotted in
Figure~\ref{fig:FluctuationSpectra} with magenta diamonds.

Another source of fluctuation is the shot noise due to faint sources.
We estimated the fluctuation
$F_{\mathrm{src}}$ produced by the sources with flux fainter than
the detection limit with the formula
$F_{\mathrm{src}} = \int S^{2} (dN/dS)\,dS$, where $S$ is the flux and
$dN/dS$ is the differential source count.
The source count functions are taken from the deep observations of the NEP with \textit{AKARI}
(\citealt{2007PASJ...59S.515W} at $7\,\micron$ and \citealt{2010A&A...514A...8P} at $15\,\micron$)
and the deep field observations with \textit{Spitzer Space Telescope}
(\citealt{2004ApJS..154...70P} at $24\,\micron$)\@.
Regarding the other wavelengths, we interpolated the source count function.
The detection limits are determined by five times the standard deviation of the
pixel values after $3\sigma$-clipping and shown in
Table~\ref{tab:IRCChannels}\@.
We accept $5\sigma$, rather than $3\sigma$, as the detection limit because of
uncertainties involved in i) the difference of source detection or removal methods
between the source-count papers and this work, ii) the difference of
calibration factors used in this work from those for point sources, and iii)
the systematic errors ($\sim 10\%$) in the calibration factors inherited from
the \textit{COBE}/DIRBE calibration.
The fluctuation spectra of the faint sources are drawn in
Figure~\ref{fig:FluctuationSpectra} with cyan dotted lines.
The spectra decrease below the scale $\sim 15\arcsec$ due to the PSF\@.
Quadratic summation of the photon-noise, the cirrus, and the shot-noise spectra are also
shown with black dashed lines.
In Table~\ref{tab:Fluctuation}, we arrange the fluctuations of dark current,
sky brightness, photon noise, Galactic cirrus, and shot noise in six
mid-infrared bands at the scale of $200\arcsec$\@.

The residual fluctuations at $200\arcsec$ after removing the photon noise, the
Galactic cirrus, and the shot noise contributions from the sky-brightness
fluctuations were calculated and are listed in Table~\ref{tab:Fluctuation}\@.
We note that the photon-noise contributions were individually subtracted from
the corresponding sky-brightness fluctuations and then the photon-noise-removed
fluctuations were averaged after rejecting the maximum and minimum values,
because it depends on the changing sky brightness.
The cirrus and the shot-noise contributions were subsequently subtracted.
Subtracting operations were done in a quadratic sense, that is, we calculated the
square-root of the difference between the squares of operands.
The errors were properly propagated.

For the MIR-S bands, the fluctuations at the scales $\lesssim 100\arcsec$ were
well described by the photon noise.
At larger scales, the slope of the spectrum becomes shallower, because the Galactic
cirrus emission surpasses the fluctuation due to the photon noise.
Owing to polycyclic aromatic hydrocarbon (PAH) features, the cirrus emission is
stronger in the short mid-infrared wavelengths than in longer wavelengths.
The contribution of the cirrus emission decreases with
the wavelength and becomes negligible in the MIR-L bands.
The photon noise dominates the sky-brightness fluctuation in the MIR-L bands at
all the scales considered in this study.
The residual fluctuation appears at scales larger than $100\arcsec$\@.
At a few arcminitues scale, the residual fluctuation is detected at all mid-infrared
bands.

We think that the residual fluctuation is related to an instrumental effect,
but not to a celestial object.
If we divide the residual fluctuations in Table~\ref{tab:Fluctuation} by the
calibration factors in Table~\ref{tab:IRCChannels} to convert the unit to ADU,
the fluctuations are about 0.7\,ADU for the MIR-S bands except for the S7 band,
and about 0.4\,ADU for the MIR-L bands.
For the S7 band, the fluctuation is the least, about $(0.26 \pm 0.24)\,\mathrm{ADU}$,
but has large error.
If we ignore the S7 band, the fluctuations at the bands of each channel
are consistent to each other in the unit of ADU\@.
Although it is difficult to point out the exact cause, we attribute the
residual fluctuations measured in this work to the instrumental effect, such as
ghost images or fluctuation of flat patterns.
In the current stage, we conclude that the residual fluctuation plus two times
error shown in Table~\ref{tab:Fluctuation} is the upper limit of the ZL
fluctuation towards the NEP at each band.
The most stringent upper limit at $200\arcsec$ scale is $\sim 0.03\%$ of the brightness at
$18\,\micron$\@.
This limit is applicable to the ZL at near-infrared wavelengths, at which the
sunlight scattered by the IPD particles is important \citep{2011ApJ...742..124M}\@.

\section{Conclusion}
\label{sec:Conclusion}

We made use of the \textit{AKARI} IRC NEP Monitor Observations to examine the
fluctuation in the sky brightness at the mid-infrared wavelengths.
The fluctuation is measured by the power-spectrum method.
After correction for the dark current,
the fluctuation in the sky brightness is retrieved at six IRC mid-infrared
bands.
At wavelengths 7, 9, and 11\,\micron, the photon noise dominates the
fluctuation spectra up to the arcminute scale, while the Galactic cirrus component
dominates above that scale.
At longer wavelengths (15, 18, and 24\,\micron), the sky-brightness fluctuation
spectra are comparable to the photon-noise spectra within a factor of two over
a scale range from 5 to 200 arcseconds.
Residual fluctuations are detected at scales larger than 100\arcsec{}
after removing the noise and cirrus
contributions from the measured sky-brightness fluctuations.
We take the smallest fluctuation, $\sim 0.03\%$ of the sky brightness at the
wavelength of $18\,\micron$, as the upper limit of the sky-brightness
fluctuation towards the NEP in the mid-infrared range.

\acknowledgments

This work is based on observations from \textit{AKARI}, a JAXA project, with the
participation of ESA\@.
This work was achieved under the auspices of Prof.~Seung~Soo~Hong.
T.M.\ and S.M.\ were supported by Grants-in-Aid from the Japan Society for the
Promotion of Science (18204018 and 21111004)\@.
This research made use of
\anchor{http://www.stsci.edu/resources/software\_hardware/pyraf}{\texttt{PyRaf}},
developed and maintained by the Space Telescope Science Institute (STScI).
All figures in the paper were prepared with
\anchor{http://matplotlib.sourceforge.net}{\texttt{matplotlib}}, an open-source
Python plotting library.

{\it Facility:} \facility{Akari (IRC)}



\clearpage


\clearpage

\clearpage

\clearpage

\clearpage

\clearpage

\clearpage

\clearpage


\begin{deluxetable}{lccccccc}
  \tablecaption{IRC Mid-infrared Channels and Photometric Bands\label{tab:IRCChannels}}
  \tablewidth{0pt}
  \tablecolumns{8}
  \tablehead{%
    \colhead{Channel} & \multicolumn{3}{c}{MIR-S} & \colhead{}
                      & \multicolumn{3}{c}{MIR-L} \\
    \cline{2-4}\cline{6-8} \\
    \colhead{Band} & \colhead{S7}  & \colhead{S9W}  & \colhead{S11} & \colhead{}
                   & \colhead{L15} & \colhead{L18W} & \colhead{L24}
  }
  \startdata
  Wavelength [\micron] & 7 & 9 & 11 & & 15 & 18 & 24 \\
  Bandwidth [\micron] & 1.75 & 4.10 & 4.12 & 
                      & 5.98 & 9.97 & 5.34 \\
  Pixel scale\tablenotemark{a} &
    \multicolumn{3}{c}{$2\farcs34 \times 2\farcs34$} & & 
    \multicolumn{3}{c}{$2\farcs51 \times 2\farcs39$} \\
  Field-of-view\tablenotemark{a} & 
    \multicolumn{3}{c}{$9\farcm1 \times 10\farcm0$} & & 
    \multicolumn{3}{c}{$10\farcm3 \times 9\farcm5$} \\
  $f_{\mathrm{diffuse}}$\tablenotemark{b} $[\mathrm{nW}\,\mathrm{m}^{-2}\,\mathrm{sr}^{-1}\,\mathrm{ADU}^{-1}]$ &
    2.93 & 1.32 & 1.39 & & 1.80 & 1.05 & 2.44 \\
  $f_{\mathrm{diffuse}} / f_{\mathrm{point}}$\tablenotemark{c} &
    0.86 & 0.89 & 0.85 & & 0.75 & 0.78 & 0.56 \\
  $5\sigma$ detection limit\tablenotemark{d} [$\mu\mathrm{Jy}$] &
    47 & 46 & 64 & & 105 & 91 & 173 \\
  \enddata
  \tablenotetext{a}{cross-scan $\times$ in-scan}
  \tablenotetext{b}{diffuse source calibration factor}
  \tablenotetext{c}{ratio of the diffuse source calibration factor to the point
  source.  $f_{\mathrm{point}}$ is taken from the 3rd column of Table~4.6.7 in
  \cite{AKARI_IRCDUM}\@.}
  \tablenotetext{d}{Five times the standard deviation after the $3\sigma$-clipping}
\end{deluxetable}

\clearpage

\begin{deluxetable}{ccrccccc}
  \tablecaption{Observation Information\label{tab:ObservationInfo}}
  \tablewidth{0pt}
  \tablecolumns{8}
  \tablehead{%
    \colhead{} & \colhead{} & \colhead{} & \multicolumn{2}{c}{MIR-S} %
                            & \colhead{} & \multicolumn{2}{c}{MIR-L} \\
    \colhead{Obs.\ ID} & \colhead{Date Time (UT)} %
                       & \colhead{$\Lambda_\oplus$\tablenotemark{a} [\degr]} %
                       & \colhead{R.A.} & \colhead{Dec.} & \colhead{} %
                       & \colhead{R.A.} & \colhead{Dec.}}
  \startdata
  5121014-001 & 2006-09-15 17:06:14 & 352.62 %
    & $17^{\mathrm{h}}55^{\mathrm{m}}23{\fs}1$ & $66{\degr}37{\arcmin}55{\farcs}4$ & %
    & $17^{\mathrm{h}}52^{\mathrm{m}}04{\fs}3$ & $66{\degr}34{\arcmin}23{\farcs}8$ \\
  5121016-001 & 2006-10-01 16:35:22 &   8.27 %
    & $17^{\mathrm{h}}55^{\mathrm{m}}22{\fs}2$ & $66{\degr}37{\arcmin}53{\farcs}4$ & %
    & $17^{\mathrm{h}}52^{\mathrm{m}}20{\fs}5$ & $66{\degr}29{\arcmin}10{\farcs}8$ \\
  5121017-001 & 2006-10-17 16:07:25 &  24.05 %
    & $17^{\mathrm{h}}55^{\mathrm{m}}21{\fs}3$ & $66{\degr}37{\arcmin}49{\farcs}2$ & %
    & $17^{\mathrm{h}}52^{\mathrm{m}}50{\fs}4$ & $66{\degr}24{\arcmin}32{\farcs}4$ \\
  5121018-001 & 2006-11-02 15:44:30 &  39.98 %
    & $17^{\mathrm{h}}55^{\mathrm{m}}20{\fs}6$ & $66{\degr}37{\arcmin}44{\farcs}5$ & %
    & $17^{\mathrm{h}}53^{\mathrm{m}}32{\fs}1$ & $66{\degr}20{\arcmin}51{\farcs}7$ \\
  5121019-001 & 2006-11-18 00:33:04 &  55.42 %
    & \nodata & \nodata & %
    & $17^{\mathrm{h}}54^{\mathrm{m}}20{\fs}3$ & $66{\degr}18{\arcmin}29{\farcs}9$ \\
  5121020-001 & 2006-12-04 00:17:29 &  71.60 %
    & \nodata & \nodata & %
    & $17^{\mathrm{h}}55^{\mathrm{m}}15{\fs}6$ & $66{\degr}17{\arcmin}29{\farcs}1$ \\
  5121021-001 & 2006-12-20 00:06:37 &  87.85 %
    & $17^{\mathrm{h}}55^{\mathrm{m}}20{\fs}3$ & $66{\degr}37{\arcmin}26{\farcs}6$ & %
    & $17^{\mathrm{h}}56^{\mathrm{m}}11{\fs}9$ & $66{\degr}18{\arcmin}03{\farcs}9$ \\
  5121022-001 & 2007-01-06 23:58:56 & 106.19 %
    & $17^{\mathrm{h}}55^{\mathrm{m}}20{\fs}8$ & $66{\degr}37{\arcmin}19{\farcs}1$ & %
    & $17^{\mathrm{h}}57^{\mathrm{m}}10{\fs}6$ & $66{\degr}20{\arcmin}31{\farcs}3$ \\
  5121023-001 & 2007-01-20 23:56:09 & 120.45 %
    & $17^{\mathrm{h}}55^{\mathrm{m}}21{\fs}4$ & $66{\degr}37{\arcmin}14{\farcs}9$ & %
    & $17^{\mathrm{h}}57^{\mathrm{m}}49{\fs}3$ & $66{\degr}23{\arcmin}38{\farcs}9$ \\
  5121024-001 & 2007-02-05 23:56:15 & 136.70 %
    & $17^{\mathrm{h}}55^{\mathrm{m}}22{\fs}4$ & $66{\degr}37{\arcmin}10{\farcs}7$ & %
    & $17^{\mathrm{h}}58^{\mathrm{m}}22{\fs}8$ & $66{\degr}28{\arcmin}13{\farcs}5$ \\
  5121025-001 & 2007-02-21 23:59:52 & 152.88 %
    & $17^{\mathrm{h}}55^{\mathrm{m}}23{\fs}3$ & $66{\degr}37{\arcmin}09{\farcs}2$ & %
    & \nodata & \nodata \\
  5121026-001 & 2007-03-10 00:06:48 & 168.93 %
    & $17^{\mathrm{h}}55^{\mathrm{m}}24{\fs}3$ & $66{\degr}37{\arcmin}08{\farcs}9$ & %
    & \nodata & \nodata \\
  \enddata
  \tablenotetext{a}{Heliocentric ecliptic longitude of the Earth at the
    observation epoch}
\end{deluxetable}

\clearpage

\begin{deluxetable}{clrcrc}
  \tablecaption{Results of Sinusoidal Fitting\label{tab:SineFitting}}
  \tablewidth{0pt}
  \tablecolumns{6}
  \tablehead{%
    \colhead{Channel} & \colhead{Band} %
      & \colhead{$a$} & \colhead{$b$} & \colhead{$c$} %
      & \colhead{$\sigma_{\mathrm{res}}$\tablenotemark{a}} \\
    & & \colhead{[$\mathrm{nW}\,\mathrm{m}^{-2}\,\mathrm{sr}^{-1}$]} %
      & \colhead{[$\degr$]} %
      & \colhead{[$\mathrm{nW}\,\mathrm{m}^{-2}\,\mathrm{sr}^{-1}$]} %
      & \colhead{[$\mathrm{nW}\,\mathrm{m}^{-2}\,\mathrm{sr}^{-1}$]}
  }
  \startdata
          & S7   & $126.93 \pm 0.12$ & $46.41 \pm 0.08$ & $1236.15 \pm 0.12$ & 4.91 (0.40\%) \\
    MIR-S & S9W  & $237.74 \pm 0.08$ & $51.96 \pm 0.03$ & $2355.60 \pm 0.09$ & 4.86 (0.21\%) \\
          & S11  & $290.45 \pm 0.09$ & $54.45 \pm 0.03$ & $3034.43 \pm 0.10$ & 9.59 (0.32\%) \\
    \hline
          & L15  & $269.70 \pm 0.13$ & $61.12 \pm 0.08$ & $3211.98 \pm 0.27$ & 5.56 (0.17\%) \\
    MIR-L & L18W & $229.03 \pm 0.10$ & $63.56 \pm 0.06$ & $2840.21 \pm 0.19$ & 4.61 (0.16\%) \\
          & L24  & $148.60 \pm 0.14$ & $69.74 \pm 0.14$ & $1984.60 \pm 0.27$ & 3.25 (0.16\%) \\
    \hline
    DIRBE & 12\,\micron{} & $284.86 \pm 0.03$ & $55.05 \pm 0.01$ & $3171.88 \pm 0.03$ & 13.73 (0.43\%) \\
          & 25\,\micron{} & $179.46 \pm 0.13$ & $69.58 \pm 0.05$ & $1891.68 \pm 0.10$ & 11.75 (0.62\%) \\
  \enddata
  \tablenotetext{a}{Standard deviation of the residuals.  Values in parentheses
    are the percentage of standard deviations with respect to the average
    brightnesses, $c$\@.}
\end{deluxetable}

\clearpage

\begin{deluxetable}{llcccccc}
  \rotate
  \tablecaption{Fluctuations at $200\arcsec$ Scale\tablenotemark{a}\label{tab:Fluctuation}}
  \tablewidth{0pt}
  \tablecolumns{8}
  \tablehead{%
    \colhead{} & \colhead{} %
      & \multicolumn{6}{c}{Fluctuation [$\mathrm{nW}\,\mathrm{m}^{-2}\,\mathrm{sr}^{-1}$]} \\
    \cline{3-8} \\
    \colhead{Channel} & \colhead{Band}
      & \colhead{Dark current} %
      & \colhead{Sky brightness} %
      & \colhead{Photon noise} %
      & \colhead{Cirrus\tablenotemark{b}}
      & \colhead{Shot noise\tablenotemark{c}}
      & \colhead{Residual\tablenotemark{d}}
  }
  \startdata
        & S7   & $0.80 \pm 0.05$ & $1.71 \pm 0.30$ & $0.71 \pm 0.01$ & $1.32 \pm 0.03$ & $0.40$          & $0.77 \pm 0.69$ ($0.174\%$) \\
  MIR-S & S9W  & $0.36 \pm 0.02$ & $1.54 \pm 0.09$ & $0.66 \pm 0.01$ & $1.01 \pm 0.02$ & $0.31$          & $0.96 \pm 0.15$ ($0.054\%$) \\
        & S11  & $0.38 \pm 0.02$ & $1.35 \pm 0.15$ & $0.76 \pm 0.01$ & $0.71 \pm 0.02$ & $0.31$          & $0.99 \pm 0.24$ ($0.049\%$) \\
  \hline
        & L15  & $0.25 \pm 0.03$ & $1.31 \pm 0.18$ & $0.88 \pm 0.01$ & $0.46 \pm 0.01$ & $0.29 \pm 0.01$ & $0.84 \pm 0.30$ ($0.045\%$)  \\
  MIR-L & L18W & $0.15 \pm 0.02$ & $0.89 \pm 0.07$ & $0.63 \pm 0.01$ & $0.37 \pm 0.01$ & $0.22 \pm 0.01$ & $0.47 \pm 0.14$ ($0.027\%$)  \\
        & L24  & $0.34 \pm 0.04$ & $1.28 \pm 0.16$ & $0.80 \pm 0.01$ & $0.26 \pm 0.01$ & $0.27 \pm 0.01$ & $1.04 \pm 0.23$ ($0.076\%$)  \\
  \enddata
  \tablenotetext{a}{The exact scale for the fluctuation measurements is
    $193\farcs05$ for the MIR-S channel and $203\farcs31$ for the MIR-L\@.}
  \tablenotetext{b}{Obtained by taking the average of the Galactic cirrus
    fluctuations between 100 and 300 arcseconds scales after rejecting the
    largest and the smallest values.  Error is the statistical error of
    average.}
  \tablenotetext{c}{Errors less than $0.01\,\mathrm{nW}\,\mathrm{m}^{-2}\,\mathrm{sr}^{-1}$
    are omitted.}
  \tablenotetext{d}{Value in parentheses is the percentage of the fluctuation
    plus two times error
    with respect to the average brightness ($c$ in Table~\ref{tab:SineFitting})
    over the seasonal variation.}
\end{deluxetable}

\clearpage

\end{document}